\documentclass{elsart}
\usepackage{graphicx}
\usepackage{amsmath}
\usepackage{amssymb}
\usepackage{setspace}

\journal{Physics Letter A}
\begin{document}
\begin{frontmatter}
\title{Quantum Message Exchange Based on Entanglement and Bell-State Measurements}
\author{Sung Soon Jang}
\and
\author{Hai-Woong Lee}
\address {Department of Physics, Korea Advanced Institute of Science and Technology Daejeon, 305-701, Korea}
\date{\today}

\begin{abstract}
We propose a scheme by which two parties can secretely and
simultaneously exchange messages. The scheme requires the two
parties to share entanglement and both to perform Bell-state
measurements. Only two out of the four Bell states are required to
be distinguished in the Bell-state measurements, and thus the
scheme is experimentally feasible using only linear optical means.
Generalizations of the scheme to high-dimensional systems and to
multipartite entanglement are considered. We show also that the
proposed scheme works even if the two parties do not possess
shared reference frames.
\end{abstract}

\end{frontmatter}

\section{Introduction}
Entanglement is an essential resource for many applications in
quantum information science such as quantum superdense
coding\cite{BW92,MWKZ96} quantum
teleportation\cite{BBCJPW93,BBMHP98,BPMEWZ97,FSBFKP98,LK01,LSPM02},
quantum cryptography\cite{E91,ECrypt00,LLCLK03}, and quantum
computing\cite{RB01,Ni04}. From an information-theoretic point of
view, two parties sharing entanglement can be regarded to have
already a certain amount of information distributed between them;
one e-bit per a shared maximally entangled pair of qubits. Thus,
for example, in superdense coding two bits of information can be
sent from one party to another by manipulating only one of two
maximally entangled qubits. In quantum teleportation a quantum
state of a qubit can be completely transferred by sending only two
bits of classical information, if the two parties, the sender and
the receiver, share a maximally entangled pair.

In this work we explore yet another situation in which two(or
more) parties can make use of entanglement they share to their
advantage. We consider a situation in which two parties, Alice and
Bob, share a maximally entangled pair A and B of qubits. Alice
makes a Bell-state measurement upon the qubit $A$ and another
qubit $\alpha$ she prepared in a state about which only she has
the information. Bob also makes a Bell-state measurement upon the
qubit $B$ and another qubit $\beta$ he prepared in a state about
which only he has the information. We are interested in the
probability for each of the sixteen possible joint measurement
outcomes, which in general depends upon the states of the qubits
$\alpha$ and $\beta$ in a way characteristic of the shared
entanglement. If Alice keeps the information on the state of qubit
$\alpha$ to herself and Bob keeps the information on the state of
qubit $\beta$ to himself, they have a partial knowledge of the
probabilities in advance that others do not. We suggest that this
advantage can be exploited to devise a method by which Alice and
Bob secretely and simultaneously exchange messages.
Generalizations of the method to higher-dimensional
systems(``qudits'') and to multipartite entanglement are also
discussed.

\section{The Basic Scheme}
Let us suppose that two parties, Alice and Bob, share a maximally
entangled pair $A$ and $B$ of qubits. The qubits A and B can be in
any of the four Bell states
\begin{equation}
|\Phi _{ij}\rangle_{AB}  = \frac{1}{\sqrt 2}\sum\limits_{q=0}^{1}
{(-1) ^{jq} } |q\rangle_A |q + i\rangle_B; i,j=0,1
\begin{array}{l}
\\ [-10mm] ,
\end{array}
\end{equation}
but for the sake of concreteness of argument, we take it as
\begin{equation}
|\Phi _{00}\rangle_{AB}  = \frac{1}{{\sqrt 2 }}\left( {\left| 0
\right\rangle_A \left| 0 \right\rangle_B  + \left| 1
\right\rangle_A \left| 1 \right\rangle_B } \right)
\end{equation}
Alice has in her possession another qubit $\alpha$ which she
prepared in state
\begin{equation}
|\psi\rangle_{\alpha} = a_0 |0 \rangle_{\alpha}+ a_1
|1\rangle_{\alpha}
\end{equation}
Similarly, Bob has in his possession yet another qubit $\beta$
which he prepared in state
\begin{equation}
|\psi\rangle_{\beta} = b_0 |0 \rangle_{\beta}+ b_1
|1\rangle_{\beta}
\end{equation}
Now Alice performs a Bell-state measurement on the pair $\alpha$
and $A$, and Bob performs a Bell-state measurement on the pair
$\beta$ and $B$. The experimental scheme is depicted schematically
in Fig.~1.

\begin{figure}
\includegraphics[width=14cm]{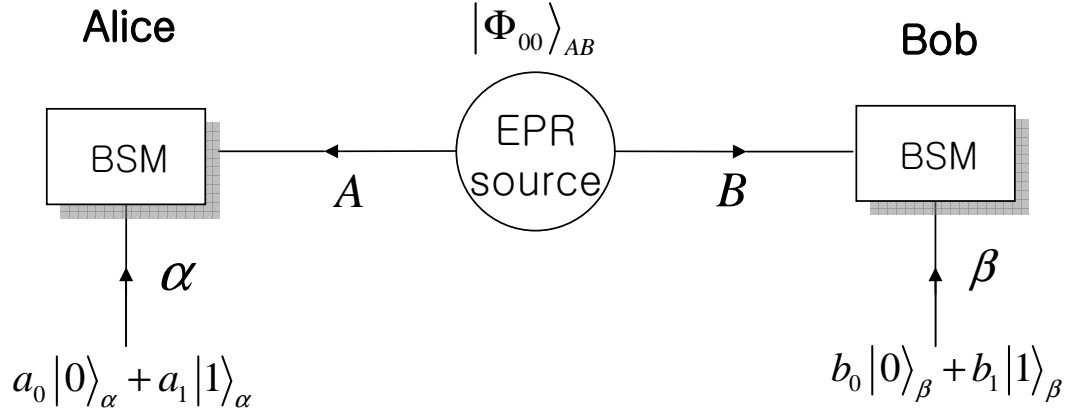} \caption{Experimental
Scheme. The EPR(Einstein-Podolsky-Rosen) source emits an entangled
pair in state $|\Phi_{00}\rangle_{AB}$. Alice performs a
Bell-state measurement on the qubit pair $\alpha$ and A, and Bob
performs a Bell-state measurement on the qubit pair $\beta$ and B.
BSM stands for Bell-state measurement.}
\end{figure}

The probability $P_{i_1 j_1 i_2 j_2} (i_1,j_1,i_2,j_2 = 0,1)$ that
Alice's Bell-state measurement yields $|\Phi_{i_1
j_1}\rangle_{\alpha A}$ and Bob's Bell-state measurement yields
$|\Phi_{i_2 j_2}\rangle_{\beta B}$ can be obtained by expanding
the total wave function $|\psi\rangle_{\alpha \beta A B} =
|\psi\rangle_\alpha |\psi\rangle_\beta |\Phi_{00}\rangle_{AB}$ as
\begin{equation}
\left| \psi  \right\rangle _{\alpha \beta AB} =
\sum\limits_{i_1,j_1,i_2,j_2 = 0}^1 {\left| {\Phi _{i_1 j_1} }
\right\rangle _{\alpha A} \left| {\Phi _{i_2 j_2} } \right\rangle
_{\beta B} V_{i_1 j_1 i_2 j_2} }
\end{equation}
A straightforward algebra yields
\begin{equation}
V_{i_1 j_1 i_2 j_2}=\frac{1}{2\sqrt 2}(-1)^{\left ( i_1 j_1 + i_2
j_2 \right )} \left [ a_{i_1} b_{i_2} + (-1)^{\left (j_1 +
j_2\right )} a_{i_1 + 1} b_{i_2 + 1} \right ],
\end{equation}
where all indices are evaluated modulo 2. The probabilities
$P_{i_1 j_1 i_2 j_2}$'s are given by $P_{i_1 j_1 i_2 j_2} =
|V_{i_1 j_1 i_2 j_2}|^2$. From Eq.(6) we immediately obtain
\begin{subequations}
\begin{eqnarray}
P_{0000}  = P_{0101}  = P_{1010} = P_{1111} = \frac{1}{8}\left |
a_0 b_0 + a_1 b_1 \right |^2 \\
P_{0001}  = P_{0100}  = P_{1011} = P_{1110} = \frac{1}{8}\left |
a_0 b_0 - a_1 b_1 \right |^2 \\
P_{0010}  = P_{0111}  = P_{1000} = P_{1101} = \frac{1}{8}\left |
a_0 b_1 + a_1 b_0 \right |^2 \\
P_{0011}  = P_{0110}  = P_{1001} = P_{1100} = \frac{1}{8}\left |
a_0 b_1 - a_1 b_0 \right |^2
\end{eqnarray}
\end{subequations}

We note that, since Alice prepared the state of qubit $\alpha$ and
thus knows what $a_0$ and $a_1$ are, and similarly since Bob
prepared the state of qubit $\beta$ and thus knows what $b_0$ and
$b_1$ are, Alice and Bob have a partial prior knowledge of the
probabilities $P_{i_1 j_1 i_2 j_2}$'s. We suggest that they can
take advantage of this knowledge to secretely exchange messages.

The scheme we propose goes as follows. We suppose that Alice and
Bob share a large number N($\gg 1$) of maximally entangled pairs
A's and B's. We further suppose that Alice has an equally large
number N of qubits $\alpha$'s, each of which she prepared in state
(3), and that Bob has N qubits $\beta$'s, each of which he
prepared in state (4). Alice keeps the information on the state of
qubits $\alpha$'s to herself and Bob keeps the information on the
state of qubits $\beta$'s to himself. Alice and Bob then perform a
series of N Bell-state measurements on each pair $\alpha$ and $A$,
and $\beta$ and $B$, respectively. They announce publicly their
measurement results only when the outcome is either $\Phi_{10}$ or
$\Phi_{11}$ (this considerably eases the burden on the Bell-state
measurements, because only these two Bell states can be
unambiguously distinguished with linear optical means
\cite{MWKZ96,LK01,LCS99}). By counting the number $N_{1 j_1 1
j_2}$ of occurrences for the joint outcome $\left |\Phi_{1 j_1}
\right \rangle_{\alpha A} \left | \Phi_{1 j_2} \right
\rangle_{\beta B}$, the probabilities $P_{1 j_1 1 j_2}$'s $(j_1 ,
j_2 = 0,1)$ can be determined experimentally as
\begin{equation}
P^{exp}_{1 j_1 1 j_2} = \frac{N_{1 j_1 1 j_2}}{N}
\end{equation}
When the four experimentally determined probabilities $P^{exp}_{1
j_1 1 j_2}$'s ($P^{exp}_{1010}$, $P^{exp}_{1111}$,
$P^{exp}_{1011}$, $P^{exp}_{1110}$) are substituted for the
probabilities $P_{1 j_1 1 j_2}$'s of Eqs. (7a) and (7b), we obtain
two equations that relate the four constants $a_0 , a_1 , b_0$ and
$b_1$. Since Alice knows the values of $a_0$ and $a_1$, there are
only two unknowns $b_0$ and $b_1$ [ the constants $b_0$ and $b_1$
are complex numbers, but they are subject to normalization and the
overall phase can be ignored] to her. Likewise, there are only two
unknowns $a_0$ and $a_1$ to Bob. To any third party, however, the
number of unknowns is four. We can thus conclude that only Alice
and Bob can completely determine the four constants $a_0, a_1,
b_0$ and $b_1$. Let us suppose that Alice and Bob have secret
messages they wish to send to each other. If they prepare their
messages in the form of two constants, the scheme described above
can be used for them to achieve a secret two-way communication. We
mention that a scheme which is different from our proposed scheme
but allows two parties to simultaneously exchange their messages
as our proposed scheme does has recently been proposed\cite{Ng04}.

\section{Efficiency}
Let us now estimate the efficiency of the scheme described in the
previous section. When a sufficiently large number $N\gg 1$ of
Bell-state measurements are made by Alice and Bob each, the number
$N_{i_1 j_1 i_2 j_2}$ of times the joint outcome $|\Phi_{i_1
j_1}\rangle_{\alpha A}|\Phi_{i_2 j_2}\rangle_{\beta B}$ is counted
lies within the range defined as\cite{Reif}
\begin{equation}
\begin{array}{l}
NP_{i_1 j_1 i_2 j_2}  - \sqrt {2NP_{i_1 j_1 i_2 j_2} \left( {1 -
P_{i_1 j_1 i_2 j_2} } \right)} \lesssim N_{i_1 j_1 i_2 j_2}^{exp}
\lesssim NP_{i_1 j_1 i_2 j_2} + \sqrt {2NP_{i_1 j_1 i_2 j_2}
\left( {1 - P_{i_1 j_1 i_2 j_2} } \right)}
\end{array}
\end{equation}
where $P_{i_1 j_1 i_2 j_2}$'s are the exact theoretical
probabilities given by Eqs.(7). Thus, the accuracy of the
experimentally determined probabilities $P^{exp}_{i_1 j_1 i_2 j_2}
= N^{exp}_{i_1 j_1 i_2 j_2}/N$ is limited by
\begin{equation}
\left | P^{exp}_{i_1 j_1 i_2 j_2} - P_{i_1 j_1 i_2 j_2} \right |
\lesssim \frac{\sqrt{2NP_{i_1 j_1 i_2 j_2} \left( {1 - P_{i_1 j_1
i_2 j_2} } \right)}}{N}\cong \frac{1}{\sqrt N}
\end{equation}
Eq.(10) is valid as long as Alice and Bob perform each of their
Bell-state measurements individually, which we assume here. [If a
collective approach is adopted, for example, if Alice performs all
N Bell-state measurements before Bob makes any of his measurements
and sends the message containing the outcome of all her
measurements to Bob, and if Bob, upon receiving Alice's message,
performs his Bell-state measurements in an appropriate collective
way, it may be possible to obtain a higher accuracy for
$P^{exp}_{i_1 j_1 i_2 j_2}$. Further research is needed on the
collective approach.] Eq.(10) indicates that, in order to obtain
$P^{exp}_{i_1 j_1 i_2 j_2}$ accurate to one decimal point, which
would allow Alice and Bob to share four real values $\cos\theta_A,
\cos\phi_A, \cos\theta_B, \cos\phi_B$ ($a_0 = \cos\theta_A, a_1 =
\sin\theta_A e^{i\phi_A}, b_0 =\cos\theta_B, a_1 = \sin\theta_B
e^{i\phi_B}$) accurate to one decimal point each, Alice and Bob
should perform $\sim 100$(perhaps a few hundred) Bell-state
measurements each. We therefore conclude that Alice and Bob gain 4
secret digits(or equivalently 13$\sim$14 secret bits) at the
expense of $\sim 100$(a few hundred) entangled pairs, i.e., the
number of secret bits gained per use of an entangled pair is
roughly 0.1 or less. The efficiency of the proposed scheme is thus
somewhat lower than that of the entanglement-based cryptographic
scheme(E91)\cite{E91}.

We note that the proposed protocol can be used for Alice and Bob
to send secretly to each other directions in space and
consequently to possess two private shared reference frames. They
need of course to store the information on their Euler angles in
the constants $a_0$, $a_1$ and $b_0$, $b_1$, respectively. Shared
reference frames are a resource for quantum communications
\cite{Enk04,RG03,BRS04}. A standard protocol\cite{RG03} to send
directions in space when Alice and Bob share entangled pairs, say
in $\left ( \Phi_{00} \right)_{AB}$, requires Alice to perform a
projection measurement in her $|0\rangle-|1\rangle$ basis on the
particle A of each entangled pair and announce publicly the
outcome of each measurement. Bob would then follow with his
projection measurement in his $|0\rangle-|1\rangle$ basis on the
corresponding particle B of each entangled pair. If a sufficiently
large number $N \gg 1$ of projection measurements are performed by
Alice and Bob, the number $N_s$ of times that Alice and Bob obtain
the same measurement outcome will be given by $\frac{N_s}{N} =
\cos^2\theta$, where $\theta$ is the angle between Alice's and
Bob's axes.[ We assume that qubits are polarized photons. If we
take spins for qubits, then $\frac{N_s}{N} =
cos^2\frac{\theta}{2}$]. Provided that Alice and Bob perform each
of their projection measurements individually, essentially the
same statistical analysis based on Eq.(9) applies here, and thus
the efficiency of our proposed protocol is of the same order of
magnitude as that of this standard entanglement-based protocol.
[If, however, we allow Bob to make collective measurements in the
standard protocol, the efficiency can be higher. See, for example,
Ref.\cite{SharingReference}.] Our protocol, however, requires both
Alice and Bob to perform Bell-state measurements, which are in
general more difficult to perform than von-Neumann projection
measurements. One advantage of our protocol is that it is
symmetric with respect to Alice and Bob and allows both Alice and
Bob to gain information, whereas in standard schemes the
information usually flows one way.

\section{Generalization to higher-dimensional systems}
We now consider a generalization of the above scheme to
higher-dimensional systems, i.e., to ``qudits''. The generalized
Bell states for a qudit can be defined as \cite{ADGJ00,E03}
\begin{equation}
|\Phi _{ij}\rangle_{AB}  = \frac{1}{\sqrt d
}\sum\limits_{q=0}^{d-1} {\omega ^{jq} } |q\rangle_A |q +
i\rangle_B;~i,j=0,1,\dots,d-1
\end{equation}
where $\omega=e^{i\frac{2 \pi}{d}}$. Let us assume that Alice and
Bob share a large number N($\gg 1$) of entangled pairs $A$'s and
$B$'s in the generalized Bell state $\left | \Phi_{00} \right
\rangle_{AB}$. Alice performs a series of N Bell-state
measurements on pairs of qudit $A$ and another qudit $\alpha$ she
prepared in state $|\psi\rangle_{\alpha} = \sum\limits_{i=0}^{d-1}
a_i |i \rangle_{\alpha}$, while Bob performs a series of N
Bell-state measurements on pairs of qudit $B$ and yet another
qudit $\beta$ he prepared in state
$|\psi\rangle_{\beta}=\sum\limits_{i=0}^{d-1}b_i|i\rangle_{\beta}$.
As in the qubit case, the total wave function
$|\psi\rangle_{\alpha\beta A B}$ can be expanded in terms of the
products of the generalized Bell states $|\Phi_{i_1
j_1}\rangle_{\alpha A} |\Phi_{i_2 j_2}\rangle_{\beta B}$ as
\begin{equation}
|\psi\rangle_{\alpha \beta A B} = \sum\limits_{i_1 , j_1 , i_2 ,
j_2 = 0}^{d-1} |\Phi_{i_1 j_1}\rangle_{\alpha A} |\Phi_{i_2
j_2}\rangle_{\beta B} V_{i_1 j_1 i_2 j_2}
\end{equation}
where $V_{i_1 j_1 i_2 j_2}$ is given by
\begin{equation}
V_{i_1 j_1 i_2 j_2}=\frac{1}{d \sqrt d}~\omega^{\left ( i_1 j_1 +
i_2 j_2 \right ) } \sum\limits_{m=0}^{d-1}\omega^{-\left (j_1 +
j_2 \right ) m} a_{m-i_1} b_{m-i_2}
\end{equation}
where all indices are now evaluated modulo d. Eq.(13) indicates
that the probabilities $P_{i_1 j_1 i_2 j_2}=\left | V_{i_1 j_1 i_2
j_2} \right|^2$ take on the same value if $i_1 - i_2$(mod d) is
the same and if $j_1 + j_2$(mod d) is the same. Thus, there are
$d^2$ different values of the probabilities $P_{i_1 j_1 i_2
j_2}$'s. Eq.(13) then provides $(d^2-1)$ independent equations
that relate the constants, $a_0, a_1,\dots,a_{d-1}$;
$b_0,b_1,\dots,b_{d-1}$. To any third party other than Alice and
Bob, the number of unknowns contained in these constants is
$(4d-4)$. There are , however, only $(2d-2)$ unknowns, as far as
Alice or Bob is concerned. By agreeing to publicly announce their
measurement results only when the outcome is among judiciously
chosen generalized Bell states, Alice and Bob can limit the number
of probabilities that can be determined experimentally in such a
way that the number of equations that relate the experimentally
determined probabilities with the constants $a_i$'s and $b_i$'s is
greater than or equal to $(2d-2)$ but less that $(4d-4)$. This
way, Alice and Bob can send secret messages in the form of
$(2d-2)$ constants to each other and as a result secretely share
$(4d-4)$ constants between them.

As an example consider the case $d=3$. If Alice and Bob announce
results of the Bell-state measurements only when they measure
either $\Phi_{00}$ or $\Phi_{21}$, they can determine
experimentally the four probabilities $P^{exp}_{0000},
P^{exp}_{2121}, P^{exp}_{0021}$ and $P^{exp}_{2100}$, which are
given by
\begin{subequations}
\begin{eqnarray}
P_{0000} & = & \frac{1}{27}\left |a_0 b_0 + a_1 b_1 + a_2 b_2
\right|^2 \\
P_{2121} & = & \frac{1}{27}\left |a_0 b_0 + a_1 b_1 \omega + a_2
b_2\omega^2 \right |^2 \\
P_{0021} & = & \frac{1}{27}\left |a_2 b_0 + a_0 b_1 \omega^2 + a_1
b_2 \omega \right |^2 \\
P_{2100} & = & \frac{1}{27}\left |a_1 b_0 \omega^2 + a_2 b_1
\omega + a_0 b_2 \right |^2
\end{eqnarray}
\end{subequations}
where $\omega=e^{i \frac{2 \pi}{3}}$. Eqs. (14a)-(14d) are
sufficient for Alice and Bob to determine their four unknowns,
allowing them to exchange messages in the form of four constants
each.

\section{Generalization to multipartite entanglement}
Another possible generalization of the proposed scheme is to the
case of multiparty communications. Let us consider the case when
N($>2$) parties share an N-qubit entangled state of
Greenberger-Horne-Zeilinger (GHZ) type \cite{GHZ} given by
\begin{equation}
|\Phi\rangle_{AB\dots N}=\frac{1}{\sqrt 2}\left ( |0\rangle_A
|0\rangle_B \dots |0\rangle_N + |1\rangle_A |1\rangle_B \dots
|1\rangle_N \right )
\end{equation}
where the letter N(also the small letter $n$ and the Greek letter
$\nu$) refers to the Nth party or Nth qubit. Each party has, in
addition to the qubit K of Eq.(15) [ the letter K (and also the
small letter k and the Greek letter $\kappa$) denotes the Kth
party or Kth qubit, where $1\leq K \leq N$], another qubit
$\kappa$ which she or he prepared in state
\begin{equation}
|\psi\rangle_\kappa = k_0 |0\rangle_\kappa + k_1 |1\rangle_\kappa
\end{equation}
Each party then performs a Bell-state measurement upon the qubits
$\kappa$ and K. The total wave function for the 2N qubits $\alpha,
\beta,\dots \nu, A, B, \dots N$ can be expanded as
\begin{equation}
|\psi\rangle_{\alpha, \beta,\dots \nu, A, B, \dots N} = \sum
\limits_{i_1, j_1, i_2, j_2, \dots, i_N, j_N = 0}^{1}\left(
\Phi_{i_1 j_1} \right )_{\alpha A}\left( \Phi_{i_2 j_2} \right
)_{\beta B} \dots \left ( \Phi_{i_N j_N} \right )_{\nu N} V_{i_1
j_1 i_2 j_2 \dots i_N j_N}
\end{equation}
A straightforward algebra yields
\begin{equation}
\begin{array}{l}
V_{i_1 j_1 i_2 j_2 \dots i_N j_N} = \\
\displaystyle{\frac{1}{(\sqrt 2)^{N+1}}}(-1)^{i_1 j_1 + i_2 j_2 +
\dots + i_N j_N } \left [ a_{i_1} b_{i_2} \dots n_{i_N} +
(-1)^{j_1 + j_2 + \dots + j_N} a_{i_1 + 1}b_{i_2 + 1} \dots n_{i_N
+1} \right ]
\end{array}
\end{equation}

In Eq.(18), the constants $n_0$ and $n_1$ define the state of the
qubit $\nu$ in which the Nth party prepared according to
\begin{equation}
|\psi\rangle_\nu = n_0 |0\rangle_\nu + n_1 |1\rangle_\nu
\end{equation}

Eq.(18) indicates that there are $2^N$ different values for the
probabilities $P_{i_1 j_1 i_2 j_2 \dots i_N j_N}$ $= \left |
V_{i_1 j_1 i_2 j_2 \dots i_N j_N} \right |^2$. To any member of
the $N$ parties sharing the entanglement of Eq.(15), the number of
unknowns is $(2N-2)$, while it is $(2N)$ to any outsider. As
before, by agreeing to publicly announce the measurement results
only when the outcome is among judiciously chosen Bell states,
each of the N parties can secretely send his message in the form
of two constants to all others of the N parties, so that the N
parties can share secretely the messages in the form of $2N$
constants.

As an example, consider the case $N=3$. Let us assume that the
three parties, Alice, Bob and Charlie agree that Charlie is the
last one to make an announcement each time, that Alice and Bob
announce the measurement results only when they obtain either
$\Phi_{10}$ or $\Phi_{11}$, and that Charlie announces his
measurement result only when both Alice and Bob announce their
measurement results and he(Charlie) obtains either $\Phi_{10}$ or
$\Phi_{11}$ or $\Phi_{00}$. The probabilities that can be
determined experimentally are then the following:
\begin{subequations}
\begin{eqnarray}
P^{exp}_{101010} &=& P^{exp}_{101111} = P^{exp}_{111110} =
P^{exp}_{111011} = \frac{1}{16}\left | a_0 b_0 c_0 + a_1 b_1 c_1
\right |^2 \\
P^{exp}_{111010} &=& P^{exp}_{101110} = P^{exp}_{101011} =
P^{exp}_{111111} = \frac{1}{16}\left | a_0 b_0 c_0 - a_1 b_1 c_1
\right |^2 \\
P^{exp}_{101100}  &=& P^{exp}_{111000} = \frac{1}{16}\left | a_0
b_0
c_1 - a_1 b_1 c_0 \right |^2 \\
P^{exp}_{101000}  &=& P^{exp}_{111100} = \frac{1}{16}\left | a_0
b_0 c_1 + a_1 b_1 c_0 \right |^2
\end{eqnarray}
\end{subequations}
Since each of Alice, Bob and Charlie has four unknowns, she or he
can use Eqs. (20a)-(20d) to solve for her or his unknowns. This
allows Alice, Bob and Charlie to share secretely six constants
among them.

If Alice, Bob and Charlie are limited to linear-optical Bell-state
measurements, they can only determine eight probabilities $P_{1
j_1 1 j_2 1 j_3}\left ( j_1, j_2, j_3 = 0, 1 \right )$ of Eqs.
(20a) and (20b) experimentally. In this situation, Alice, Bob and
Charlie each needs to announce publicly one of the constants, say
$a_0, b_0$ and $c_0$. Each of Alice, Bob and Charlie then has two
unknowns for which Eqs.(20a) and (20b) provide sufficient
information. In this case, however, the number of constants that
the three parties can secretely share is reduced to three.

\section{Case of no shared reference frames}
So far, we have assumed that Alice and Bob have exactly the same
basis for the states $|0\rangle$ and $|1\rangle$. Thus, if the
qubits we consider are polarized photons or spins, we assume that
Alice and Bob share a spatial reference frame. As far as our
proposed protocol is concerned, this shared reference frame does
not need to be private, because the privacy of the protocol does
not depend upon the privacy of the reference frame. The shared
reference frame can, for example, be a specific direction with
respect to a fixed star.

\begin{figure}
\center
\includegraphics[width=5cm]{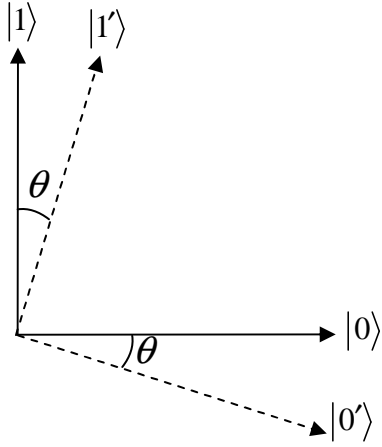} \caption{Alice's reference
frame ($|1\rangle$ and $|0\rangle$) and Bob's reference frame
($|1'\rangle$ and $|0'\rangle$).}
\end{figure}

It may happen, however, that Alice and Bob, for reasons of better
security, want to use reference frames of their own choice which
may not coincide, or that their reference frames are inadvertently
misaligned. As show below, our proposed protocol still works, even
if Alice's reference frame does not coincide with Bob's. Let us
suppose that Bob's axis makes an angle $\theta$ with respect to
Alice's, as shown in Fig.~2. The initial state for the four
particles $\alpha,\beta,A$ and $B$ are now written as
\begin{equation}
\left|\Psi\right\rangle_{\alpha\beta A B} = \left( {a_0 \left| 0
\right\rangle _\alpha + a_1 \left| 1 \right\rangle _\alpha  }
\right) \frac{1}{\sqrt 2} \left( |0\rangle_A |0\rangle_B +
|1\rangle_A |1\rangle_B \right) \left( {b_0 \left| {0'}
\right\rangle _\beta + b_1 \left| {1'} \right\rangle _\beta }
\right)
\end{equation}
where$|0\rangle$ and $|1\rangle$ denote Alice's basis states and
$|0'\rangle$ and $|1'\rangle$ Bob's basis states. We assume that
Alice prepares the entangled pair $A$ and $B$ in
$|\Phi_{00}\rangle_{AB}$, keeps $A$ and sends $B$ to Bob. Now
Alice performs her Bell state measurement on the pair $\alpha A$
in her $|0\rangle-|1\rangle$ basis, whereas Bob performs his Bell
state measurement on the pair $\beta B$ in his
$|0'\rangle-|1'\rangle$ basis. We thus need to express
$|0\rangle_B$ and $|1\rangle_B$ in terms of $|0'\rangle_B$ and
$|1'\rangle_B$, and expand the wave function $|\Psi\rangle_{\alpha
\beta A B}$ as
\begin{equation}
\left|\Psi\right\rangle_{\alpha \beta A B} = \sum\limits_{i_1 ,j_1
,i_2,j_2=0}^1 \left| {\Phi _{i_1 j_1} } \right\rangle_{\alpha A}
\left| {\Phi '_{i_2 j_2} } \right \rangle_{\beta B} V_{i_1 j_1 i_2
j_2}
\end{equation}
where $|\Phi'_{ij}\rangle$ refers to Bell states of Eq. (1)
defined  in terms of Bob's basis $|0'\rangle$ and $|1'\rangle$. A
straightforward algebra yields, for the probabilities $P_{i_1 j_1
i_2 j_2}=\left| V_{i_1 j_1 i_2 j_2} \right|^2$,
\begin{subequations}
\begin{equation}
P_{0000} = P_{1111}  = \frac{1}{8}\left| {a_0 b_0 \cos \theta -
a_1 b_0 \sin \theta  + a_0 b_1 \sin \theta  + a_1 b_1 \cos \theta
} \right|^2 \equiv \frac{1}{2}P_1
\end{equation}
\begin{equation}
P_{0001} = P_{1110} = \frac{1}{8}\left| {a_0 b_0 \cos \theta - a_1
b_0 \sin \theta  - a_0 b_1 \sin \theta  - a_1 b_1 \cos \theta }
\right|^2 \equiv \frac{1}{2}P_2
\end{equation}
\begin{equation}
P_{0010} = P_{1101}  = \frac{1}{8}\left| {a_0 b_0 \sin \theta +
a_1 b_0 \cos \theta  + a_0 b_1 \cos \theta  - a_1 b_1 \sin \theta
} \right|^2 \equiv \frac{1}{2}P_3
\end{equation}
\begin{equation}
P_{0011} = P_{1100}  = \frac{1}{8}\left| {a_0 b_0 \sin \theta +
a_1 b_0 \cos \theta  - a_0 b_1 \cos \theta  + a_1 b_1 \sin \theta
} \right|^2 \equiv \frac{1}{2}P_4
\end{equation}
\begin{equation}
P_{0100} = P_{1011}  = \frac{1}{8}\left| {a_0 b_0 \cos \theta +
a_1 b_0 \sin \theta  + a_0 b_1 \sin \theta  - a_1 b_1 \cos \theta
} \right|^2 \equiv \frac{1}{2}P_5
\end{equation}
\begin{equation}
P_{0101} = P_{1010}  = \frac{1}{8}\left| {a_0 b_0 \cos \theta +
a_1 b_0 \sin \theta  - a_0 b_1 \sin \theta  + a_1 b_1 \cos \theta
} \right|^2 \equiv \frac{1}{2}P_6
\end{equation}
\begin{equation}
P_{0110} = P_{1001}  = \frac{1}{8}\left| {a_0 b_0 \sin \theta -
a_1 b_0 \cos \theta  + a_0 b_1 \cos \theta  + a_1 b_1 \sin \theta
} \right|^2 \equiv \frac{1}{2}P_7
\end{equation}
\begin{equation}
P_{0111} = P_{1000}  = \frac{1}{8}\left| {a_0 b_0 \sin \theta -
a_1 b_0 \cos \theta  - a_0 b_1 \cos \theta  - a_1 b_1 \sin \theta
} \right|^2 \equiv \frac{1}{2}P_8
\end{equation}
\end{subequations}
where the probabilities are further restricted by the identities
\begin{subequations}
\begin{equation}
P_1 + P_2 + P_3 + P_4 = \frac{1}{2}
\end{equation}
\begin{equation}
P_2 + P_3 = P_5 + P_8
\end{equation}
\begin{equation}
P_1 + P_4 = P_6 + P_7
\end{equation}
\end{subequations}
Comparison of Eqs.(23) with Eqs.(7) indicates that the
misalignment of Bob's axis with respect to Alice's axis partly
breaks degeneracies among the probabilities. For example,
$P_{1111}$ is no longer equal to $P_{1010}$, and $P_{1110}$ is no
longer equal to $P_{1011}$. The difference between these
probabilities can thus be considered as a measure of the
misalignment.

Our scheme for quantum message exchange can proceed exactly as
before. We let Alice and Bob announce publicly their measurement
results only when the outcome is either $\Phi_{10}$ or
$\Phi_{11}$. The four experimentally determined probabilities
$P^{exp}_{1010}, P^{exp}_{1111}, P^{exp}_{1011}$, and
$P^{exp}_{1110}$ then provide four equations, Eqs. (23a), (23b),
(23e) and (23f), that relate the five constants; $a_0, a_1, b_0,
b_1$ and the angle $\theta$. To Alice(Bob) there are three
unknowns $b_0, b_1 (a_0, a_1)$ and $\theta$, whereas to any third
party the number of unknowns is five. Only Alice and Bob can thus
determine the five constants $a_0, a_1, b_0, b_1$ and $\theta$.
Our scheme thus allows Alice and Bob not only to secretely share
the four constants but also to determine the angle between their
reference frames.

\section{Summary and discussion}
We have analyzed a situation in which each of two (or more)
parties sharing entanglement performs a Bell-state measurement
upon the entangled particle in his (or her) possession and another
particle he(or she) prepared in a specific state. The probability
for a joint measurement outcome corresponding to a given
combination of the Bell states depends critically upon the states
of the particles involved. Taking advantage of the fact that each
person belonging to the parties sharing entanglement and only
he(or she) knows the state of the particle he(or she) prepared, we
suggest a scheme by which two(or more) parties sharing
entanglement can secretely and simultaneously exchange messages.
For the case of two parties sharing entangled qubits, the scheme
requires Bell-state measurements that distinguish only two out of
the four Bell states, which can be accomplished using only linear
optical devices. The scheme thus provides an experimentally
feasible means of two-way communication.

We should emphasize that, although it may not be apparent at first
sight, entanglement plays a critical role in the proposed scheme.
It is through entanglement that the joint probabilities appear in
an ``entangled'' form as give in Eqs.(7) and that the separate
probabilities for either Alice or Bob to obtain any arbitrary Bell
state are evenly distributed regardless of which Bell state we
consider. Information on the constants $a_0,a_1,b_0$ and $b_1$ can
be obtained only by looking at the joint probabilities. On the
other hand, if the qubits A and B were not entangled, Alice's
Bell-state measurements would be completely independent of Bob's
Bell-state measurements, and information on the constants $a_0$
and $a_1$ ($b_0$ and $b_1$) would be obtained by looking only at
the results of Alice's (Bob's) Bell-state measurements. Bob(Alice)
would need as much information as any third party in order to
determine $a_0$ and $a_1$($b_0$ and $b_1$) from the results of
Alice's (Bob's) Bell-state measurements. The parties sharing
entanglement have advantages only because the joint probabilities
for Alice's and Bob's Bell-state measurements are ``entangled''.
Of course, the maximal advantage is provided by the maximal
entanglement which we have assumed. In general, as the degree of
shared entanglement is decreased, the joint probabilities exhibit
less degree of entanglement, and as a result the parties sharing
entanglement has less degree of advantage over a third party.

\ack This research was supported by a Grant from the Korea Science
and Engineering Foundation(KOSEF) through Korea-China
International Cooperative Research Program and from the Ministry
of Science and Technology(MOST) of Korea. The authors thank
Professor Gui Lu Long of Tsinghua university of China for helpful
discussions.


\begin{thebibliography}{99}
  \renewcommand{\baselinestretch}{1}
  \selectfont
\bibitem{BW92}
  C. H. Bennett and S. J. Wiesner, Phys. Rev. Lett.
  \textbf{69} (1992) 2881.
\bibitem{MWKZ96}
  K.Mattle, H.Weinfurter, P.G.Kwiat, and A. Zeilinger,
  Phys. Rev. Lett. \textbf{76} (1996) 4656.
\bibitem{BBCJPW93}
  C.H.Benett, G. Brassard, C. Crepeau, R. Jozsa, A. Peres, and W.
  K. Wootters, Phys. Rev. Lett. \textbf{70} (1993) 1895.
\bibitem{BBMHP98}
  D. Boschi, S. Branca, F. De Martini, L. Hardy, and S.
  Popescu, Phys. Rev. Lett. \textbf{80} (1998) 1121.
\bibitem{BPMEWZ97}
  D. Bouwmeester, J.W. Pan, K. Mattle, M. Eibl, H.
  Weinfurter, and A. Zeilinger, Nature (London) \textbf{390} (1997)
  575.
\bibitem{FSBFKP98}
  A. Furusawa, J. L. S{\o}rensen, S.L.Braustein, C. A. Fuchs, H.
  J. Kimble, and E. S. Polzik, Science \textbf{282} (1998) 706.
\bibitem{LK01}
  H.W.Lee and J. Kim, Phys. Rev. A \textbf{63} (2001) 012305.
\bibitem{LSPM02}
  E. Lombardi, F. Sciarrino, S. Popescu, and F. De Martini, Phys.
  Rev. Lett. \textbf{88} (2002) 070402.
\bibitem{E91}
  A. K. Ekert, Phys. Rev. Lett. \textbf{67} (1991) 661.
\bibitem{ECrypt00}
  T. Jennewein, C. Simon, G. Weihs, H. Weinfurter, and A.
  Zeilinger, Phys. Rev. Lett. \textbf{84} (2000) 4729;

  D. S. Naik, C. G. Peterson, A. G. White, A. J. Berglund, and P.
  G. Kwiat, Phys. Rev. Lett. \textbf{84} (2000) 4733;

  W. Tittel, J. Brendel, H. Zbinden, and N. Gisin, Phys. Rev. Lett. \textbf{84}
  (2000) 4737.
\bibitem{LLCLK03}
  J.W.Lee, E.K.Lee, Y.W.Chung, H.W.Lee, and J.Kim, Phys. Rev. A
  \textbf{68} (2003) 012324.
\bibitem{RB01}
  R. Raussendorf and H.J.Briegel, Phys. Rev. Lett. \textbf{86} (2001)
  5188.
\bibitem{Ni04}
  M.A.Nielsen, quant-ph/0402005.
\bibitem{LCS99}
  N.L\"{u}tkenhaus, J.Calsamiglia, and S.A.Suominen, Phys. Rev. A
  \textbf{59} (1999) 3295.
\bibitem{Ng04}
  B.A.Nguyen, Phys. Lett. A \textbf{328} (2004) 6.
\bibitem{Reif}
  See, for example, F. Reif, Fundamentals of Statistical and
  Thermal Physics (McGraw-Hill, New York, 1965), Ch. 1.
\bibitem{Enk04}
  S.J.van Enk, quant-ph/0410083
\bibitem{RG03}
  T.Rudolph and L.Grover, Phys.Rev.Lett. \textbf{91},
  217905(2003).
\bibitem{BRS04}
  S.D.Bartlett, T.Rudolph, and R.W.Spekkens, Phys.Rev.A
  \textbf{70}, 032307 (2004).
\bibitem{SharingReference}
  S.Massar and S.Popescu, Phys.Rev.Lett. \textbf{74}, 1259(1995);

  R.Derka, V.Buzek, and A.K.Ekert, Phys.Rev.Lett. \textbf{80},
  1571 (1998);

  N.Gisin and S.Popescu, Phys.Rev.Lett. \textbf{83}, 432 (1999);

  S.Massar, Phys.Rev.A \textbf{62}, 040101(R) (2000);

  A.Peres and P.F.Scudo, Phys.Rev.Lett. \textbf{86}, 4160 (2001).
\bibitem{ADGJ00}
  G.Alber, A.Delgado, N.Gisin, and I.Jex, quant-ph/0008022.
\bibitem{E03}
  S.J. van Enk, Phys. Rev. Lett. \textbf{91} (2003) 017902.
\bibitem{GHZ}
  D.M.Greenberger, M.A.Horne, A.Shimony, and A.Zeilinger,
  Amer.J.Phys. \textbf{58} (1990) 1131;

  N.D.Mermin, Phys. Today \textbf{43} (1990) 9.
\end{thebibliography}
\end{document}